\documentclass[ALICE,manyauthors]{cernphprep}
\usepackage[comma,square,numbers,sort&compress]{natbib}
\usepackage{hyperref}
\usepackage{lineno}
\usepackage{inputenc}
\usepackage{fontenc}
\usepackage{hyperref}
\usepackage{graphicx}  
\usepackage{dcolumn}   
\usepackage{bm}        
\usepackage{amssymb}   
\usepackage{amsfonts}
\usepackage{graphics}
\usepackage{grffile}   
\usepackage{epsfig}
\usepackage{units}
\usepackage[usenames]{color}
\usepackage[normalem]{ulem} 
\usepackage{lineno}
\usepackage{color}
\usepackage{subfigure}

\newcommand{ \be }{\begin{linenomath*}\begin{eqnarray}}
\newcommand{ \ee }{\end{eqnarray}\end{linenomath*}}
\newcommand{ \la }{\langle}

\newcommand{ \ra }{\rangle}

\newcommand{ \mean }[1]{\left\langle #1 \right\rangle}

\newcommand{ \pvecOne}{\vec{p}_1}
\newcommand{ \pvecTwo}{\vec{p}_2}
\newcommand{ \pt }{p_{\rm T}}
\newcommand{ \ptOne}{p_{\rm T,1}}
\newcommand{ \ptTwo}{p_{\rm T,2}}
\newcommand{ \pti}{p_{\rm T,i}}

\newcommand{ \dpt }{dp_{\rm T}}
\newcommand{ \dptOne}{dp_{\rm T,1}}
\newcommand{ \dptTwo}{dp_{\rm T,2}}

\newcommand{ \Dpt }{\Delta p_{\rm T}}

\newcommand{ \Dphi }{\Delta \varphi}

\newcommand{ \Deta }{\Delta \eta}

\newcommand{ \DetaDphi }{\left(  \Delta \eta, \Delta \varphi \right)}

\newcommand{ \rhoTwo}{\rho_2}

\newcommand{\rhoDptDpt}{\la \Delta p_{\rm T}  \Delta p_{\rm T} \ra }

\definecolor{dgreen}{cmyk}{1.,0.,1.,0.2}        
\definecolor{orange}{cmyk}{0.,0.353,1.,0.}    

\begin{document}%

\begin{titlepage}
\PHyear{2017}
\PHnumber{021}      
\PHdate{30 Jan}

\title{Flow dominance and factorization of transverse momentum
  correlations in Pb--Pb collisions at the LHC} \ShortTitle{Transverse
  momentum correlations}

\Collaboration{ALICE Collaboration\thanks{See
    Appendix~\ref{app:collab} for the list of collaboration members}}
\ShortAuthor{ALICE Collaboration} 

\begin{abstract}
  We present the first measurement of the two-particle
  transverse momentum differential correlation function,
  $P_2\equiv\la \Delta p_{\rm T}  \Delta p_{\rm T} \ra /\la p_{\rm T} \ra^2$, in Pb--Pb collisions 
  at $\sqrt{s_{_{\rm NN}}} =$ 2.76 TeV. Results for $P_2$ are reported as a function of
  relative pseudorapidity ($\Delta \eta$) and azimuthal angle
  ($\Delta \varphi$) between two particles for different collision centralities. 
  The $\Delta \phi$ dependence is found
  to be largely independent of $\Delta \eta$ for $|\Delta \eta| \geq$ 0.9. 
  In 5\% most central Pb--Pb
  collisions, the two-particle transverse momentum correlation
  function exhibits a clear double-hump structure around
  $\Delta \varphi = \pi$ (i.e., on the away side), 
  which is not observed in number correlations in the same centrality range, and
  thus provides an indication of the dominance of triangular flow in
  this collision centrality. Fourier decompositions of $P_2$, studied as a
  function of collision centrality, show that correlations at 
  $|\Delta \eta| \geq$ 0.9 can be well reproduced by a flow ansatz based on the notion
  that measured transverse momentum correlations are strictly determined by the
  collective motion of the system. 
\end{abstract}

\end{titlepage}

\setcounter{page}{2} 
Measurements of particle production and their correlations in heavy-ion 
collisions at the Relativistic Heavy Ion Collider (RHIC) and the
Large Hadron Collider (LHC) have provided very compelling evidence
that the produced matter is characterized by extremely high
temperatures and energy densities consistent with a deconfined, but
strongly interacting quark-gluon plasma (sQGP). Evidence for the production 
of the sQGP is provided by observations of 
large suppression of particle production at momenta $\pt \gtrsim 3$~GeV/{\it{c}} relative
to that observed in pp collisions, strong suppression of
away-side particles observed in two-particle number
correlations, as well as by anisotropic flow studies (anisotropies
in particle azimuthal distributions relative to the reaction
plane defined by the beam axis and a line connecting the centers of colliding nuclei) 
~\cite{RHICqgp1,RHICqgp2,RHICqgp3, RHICqgp4, LHCqgp1, LHCqgp2, LHCqgp2A, 
  LHCqgp3, LHCqgp4, LHCqgp5, LHCqgp6}. The comparison of measured flow 
coefficients, $v_{n}$, with predictions from hydrodynamical models indicate that the 
sQGP has a vanishingly small shear viscosity over
entropy density ratio~\cite{Heinz:2013th}. Furthermore, the observation of an approximate number of constituent quark scaling of 
flow coefficients in the $2 < p_T < 4$ GeV/$c$  range, suggested as a signature of a deconfined medium~\cite{Molnar}, was reported by RHIC and LHC experiments~\cite{STAR-PRC75-2007-054906,Abelev:2014pua}.
These results imply that the two-particle number correlations observed 
in the region of low $\pt$ ($< 2$ GeV/{\it{c}}), corresponding to the bulk of 
particle production, are largely determined by anisotropic flow. Such flow
dominance is manifested, in particular, by an approximate factorization
of the measured flow coefficients, $V_{n\Delta}(\eta_1,\ptOne,\eta_2,\ptTwo)
= \mean{\cos(n\Delta\varphi)} = \mean{v_n(\eta_1,\ptOne)
  v_n(\eta_2,\ptTwo)}$, observed for pairs of particles at relative
pseudorapidity $\Delta\eta>0.8$, in different transverse momentum bins
up to $\pt \approx 3-5$~GeV/{\it{c}}~\cite{ALICE-harmonic}.

Two-particle transverse momentum
correlations~\cite{monica,Voloshin:1999yf,meanpt1,Adamova:2008sx,Abelev:2014ckr}
provide additional insights into the dynamics of multiparticle
production and can be used to further examine the flow dominance of
two-particle correlation functions. One expects, in particular, that
in the presence of anisotropic flow, the differential
transverse momentum correlator $\la \Dpt \Dpt\ra$ 
should feature azimuthal Fourier decomposition coefficients
calculable with a simple formula, hereafter called the {\it flow ansatz},
in terms of the regular and $\pt$ weighted flow
coefficients~\cite{monica}. Such a simple relation, discussed in more
details below, is not expected for particle production arising from
processes not related to the common symmetry plane, known as
non-flow, such as jets or resonance decays. An agreement
between the Fourier coefficients of the $\la \Dpt \Dpt \ra$ correlator and those calculated with the flow ansatz
should thus provide additional evidence of the dominance of collective
flow effects.

In this Letter, we present the first measurements of the differential transverse
momentum correlations 
in Pb--Pb collisions at $\sqrt{s_{\rm NN}} =$ 2.76
TeV in terms of the dimensionless correlator 
$P_2$ defined as
\be 
\label{eq:Rpt}
 P_{2} = \frac{\rhoDptDpt
  \DetaDphi}{\la \pt \ra^2} = \frac{1}{\la \pt \ra^2}
\frac{\int_{p_{\rm T,min}}^{p_{\rm T,max}} ~ {\rhoTwo(\pvecOne,\pvecTwo)\, ~
    \Delta \ptOne \Delta \ptTwo ~~ \dptOne \dptTwo}}
     {\int_{p_{\rm T,min}}^{p_{\rm T,max}} {\rhoTwo(\pvecOne,\pvecTwo) ~~
         \dptOne \dptTwo}},
\ee
where $\Delta\pti=\pti-\la \pt  \ra$, with $\la
\pt  \ra = \int \rho_1 ~ \pt  ~ d\pt ~ / \int \rho_1 ~
d\pt $, the inclusive average transverse momentum of
particles observed in the $p_{\rm T,min} \le \pt  \le p_{\rm T,max}$ range.
 The quantities 
$\rho_1$ and $\rho_2$ represent  single and two-particle densities,
respectively.
For particle correlations induced strictly by anisotropic emission relative to
the reaction plane, the Fourier coefficients of
$P_2$, $v_{n}[P_2]$, should be determined by regular and the 
$\pt $ weighted
flow coefficients defined according to the following 
{\it flow ansatz}~\cite{monica}:
\begin{eqnarray}\label{eq:ansatz}
v_{n}[P_2] \cong v_{n}^{\pt }/\langle \pt 
\rangle - v_{n},
\end{eqnarray}
where $v_n$ and $v_{n}^{\pt } =$ 
$\int \rho_1 v_n(\pt ) \pt  \dpt /\int \rho_1 \dpt $ 
are the regular, and $\pt $ weighted
coefficients, respectively~\cite{monica,Voloshin:2008dg}.
Thus, we shall compare the 
Fourier coefficients of the $P_2$ correlator to values expected from
this ansatz based on coefficients $v_n$ and $v_{n}^{\pt }$ measured
with traditional flow methods, e.g., the scalar product method~\cite{Voloshin:2008dg}. 

This study is based on an analysis of a $14 \times10^6$ events subset of a sample of minimum bias
(MB) trigger events recorded with the ALICE detector during the LHC
Run $1$ in $2010$. Detailed descriptions of the ALICE detector, its
subsystems, and their respective performance have been reported 
in~\cite{ALICE1, ALICE2, ALICE3, Pileup}. For this study, the Inner
Tracking System (ITS) and the Time Projection Chamber (TPC) were used
to reconstruct charged-particle tracks, while the V0 detector and the
Silicon Pixel Detector (SPD) formed the basis of the online minimum
bias trigger used to acquire the data, as described in ~\cite{LHCqgp1,LHCqgp2}. 

The ALICE solenoidal magnet was operated with a
field of 0.5 $T$ with both positive and negative polarities.  Events
included in this analysis were required to have a single reconstructed
primary vertex within 10 cm of the nominal interaction point along the
beam axis, hereafter taken to be the $z$-axis.  The fraction of pile-up
events in the analysis sample is found to be negligible after applying dedicated
pile-up removal criteria~\cite{Pileup}.

Correlation functions reported in this article are based on
charged-particle tracks measured in the pseudorapidity range $|\eta| <
1.0$ and with full azimuthal coverage $0\le \varphi < 2\pi$. The
analysis was limited to particles produced with $0.2 < \pt  <$
2.0 GeV/{\it{c}} corresponding largely to particles emerging from the bulk of
the matter. Only tracks with a minimum of 70 reconstructed space
points in the TPC, out of a maximum of 159, were included in
the analysis.  
Contributions from photon conversions into
$e^{+}e^{-}$ pairs were suppressed based on an electron rejection
criterion relying on the truncated average of the specific
ionization energy loss $\langle dE/dx \rangle$ measured in
the TPC. Tracks with $\langle dE/dx \rangle$ lying within
3$\sigma_{dE/dx}$ of the Bethe-Bloch parametrization of the $dE/dx$
expectation value for electrons and at least 3$\sigma_{dE/dx}$ away
from the relevant parameterizations for $\pi$, $K$, $p$ were removed.
In addition, suppression of the contamination from secondary particles
originating from weak decays and from interaction of particles with
the detector material was accomplished by imposing upper limits
of 3.2 cm  and 2.4 cm  (RMS $\sim$ 0.36 cm) for the distance of closest approach (DCA) of a
track to the reconstructed vertex in the longitudinal (${\rm DCA}_{z}$) and
radial (${\rm DCA}_{xy}$) directions, respectively.  
These criteria lead to a reconstruction efficiency of about 80\%
for primary particles and contamination from secondaries of about 5\%
at $p_{\rm{T}}$ = 1 GeV/{\it{c}}~\cite{ALICEeff}.  
No filters were used to
suppress like-sign (LS) particle correlations resulting from Hanbury Brown
and Twiss (HBT) effects, which produce a strong and narrow peak centered at $\Delta\eta,\Delta\varphi=0$ in LS correlation functions. Corrections for single track losses were
carried out using the weight technique described in~\cite{method_prc}
with weights calculated separately for positively and negatively
charged tracks, positive and negative 
solenoidal magnetic fields, and with 40 vertex position bins in the
fiducial range $|z|\le 10$ cm.  Pair inefficiencies associated with
track merging or crossing (e.g. two tracks being partly or entirely
reconstructed as a single track) within the TPC were corrected for
based on track charge and momentum ordering
techniques~\cite{monica_star}.  The $P_2$ correlators were measured separately
for charge pair combinations $++$, $+-$, and $--$ and were combined with
equal weights to produce the charge independent correlation functions
reported in this article.

Systematic uncertainties were investigated by repeating the analysis
 for different operational and analysis conditions including two  solenoidal magnetic field polarities, different event and track selection criteria, as well as
different track reconstruction methods. Track selection criteria, most particularly the maximum value of distance of closest approach to the primary vertex,   dominate systematic effects.   The systematic uncertainties 
assigned to the measurements of $v_{n}$ coefficients are the quadratic 
sums of individual 
contributions and range from 4\% in central 0 -- 10\% collisions to 14\% in
peripheral 70 -- 80\% collisions. 

\begin{figure}
\includegraphics[scale=0.43]{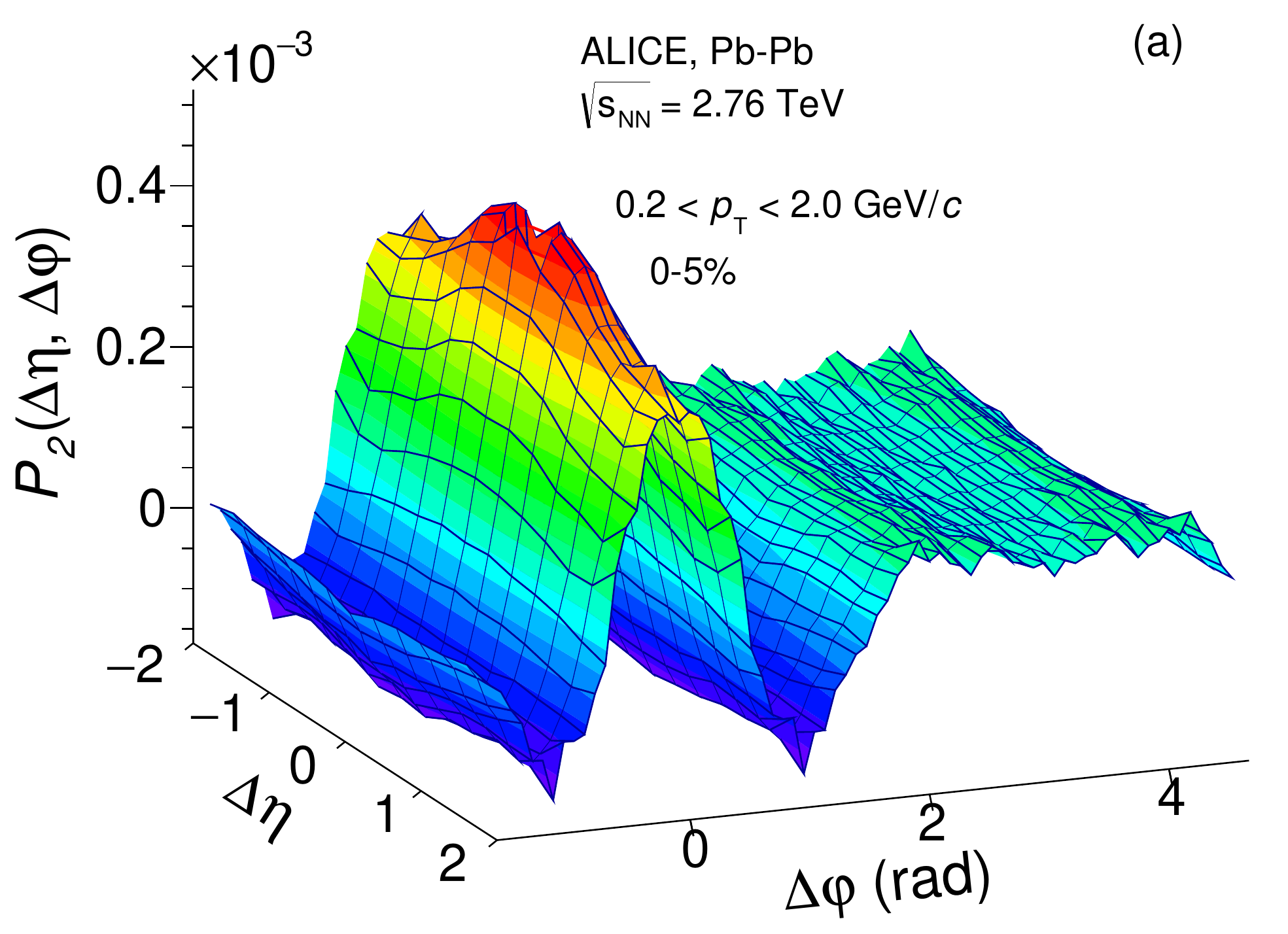}
\includegraphics[scale=0.37]{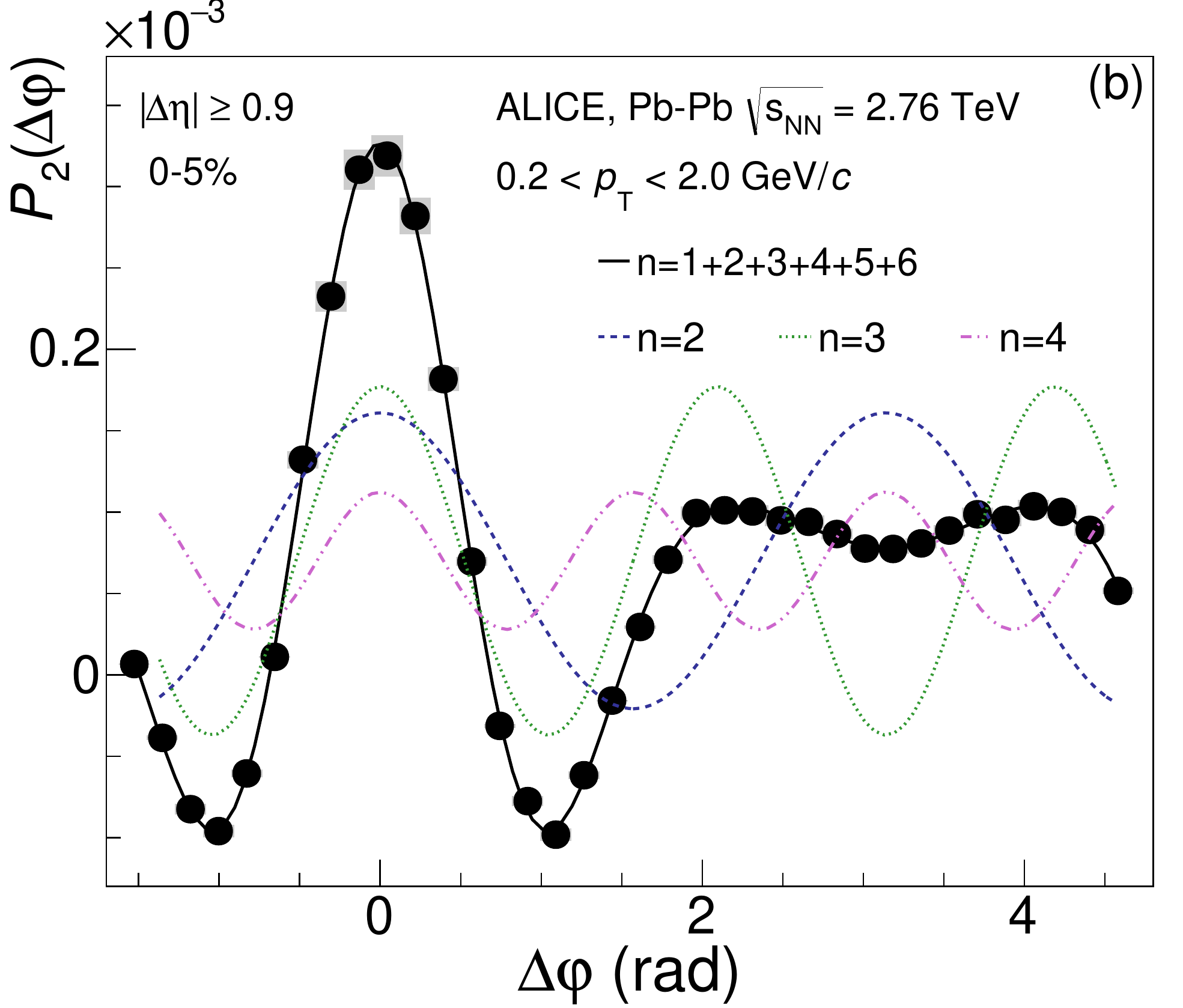}
\caption{(Color online) (a) $P_2 (\Deta,\Dphi) $ in the  5\% most central
  Pb--Pb collisions. The region $|\Deta| < 0.15$ and
  $|\Delta\varphi| < 0.13$ rad, where the weight technique used in
  this analysis does not provide a reliable efficiency correction, is excluded. 
  (b) $P_2 (\Dphi)$ for $|\Delta\eta| \geq 0.9$. 
  Systematic errors are shown as gray boxes. Note the statistically significant
  dip at $\Delta\varphi \sim \pi$. }
\label{fig:Figure1}
\end{figure}

Figure~\ref{fig:Figure1} (a) presents the correlator $P_2$ measured
as a function of $\Delta\eta$ and $\Delta\varphi$ in the  5\% most central Pb--Pb
collisions. The central range around $\Delta\eta\sim 0$ and $\Delta\varphi\sim 0
(\rm rad)$ is left under-corrected by the weight correction
procedure mainly due to track merging effects. It is thus not considered in this analysis. \footnote{The central range around $\Delta\eta\sim 0$ and $\Delta\varphi\sim 0
(\rm rad)$  is considered, however, in a related ALICE analysis carried out with a mixed event technique~\cite{janFiete}. }
The correlator $P_2$ features a prominent
near-side ridge centered at $\Delta\varphi=0$, extending across the full
pseudorapidity range of the measurement. It also features two distinct
away-side humps at $|\Dphi - \pi| \approx 60^{\circ}$ separated by a
weak dip centered at $\Dphi=\pi$ and also extending across the full
pseudorapidity range of the acceptance. Such an away-side correlation
feature, which indicates the presence of a strong third harmonic, was previously reported in ultra-central (0--2\%) Pb--Pb
collisions at the LHC~\cite{ALICE-harmonic, CMS-harmonic, ATLAS-harmonic} 
as well as in central Au--Au
collisions at RHIC but for the latter case only after the subtraction of a correlated
component whose shape was exclusively attributed to elliptic
flow~\cite{rhicv3_a, rhicv3_b, rhicv3_c}.
  
To further study the azimuthal angle dependence of transverse momentum
correlations, projections of the measured $P_2$ correlation function
are fitted with an unconstrained 6-th order Fourier decomposition in
$\Dphi$ according to $F(\Dphi) = b_0 + 2 \sum_{n=1}^{6} b_n
~\cos(n\Dphi)$, as illustrated in Fig.~\ref{fig:Figure1} (b). 
We verified that higher-order contributions, with $n>6$, do not
significantly improve the fits for $|\Deta|\geq$ 0.9.  Coefficients $b_5, b_6$  feature large relative
errors and are thus not reported in this paper.
The double-hump at $|\Dphi-\pi| \approx 60^{\circ}$ implies the 
presence of a strong third
harmonic, $v_3$, in the Fourier decompositions of the correlation
functions. 
 The large $v_{3}$ likely originates  from
fluctuations in the initial density profile of colliding nuclei~\cite{AlversRoland}.

The flow coefficients obtained from two-particle transverse momentum
correlations, $v_n[P_2]$,  calculated according to 
$v_n = \sqrt{b_n/(b_{0}+1) }$, are  plotted
in Fig.~2 as a function of centrality for central (0 -- 5\%) up to peripheral collisions
(70 -- 80\%). The $v_n[P_2]$ coefficients exhibit a collision
centrality dependence qualitatively similar to that of regular flow
coefficients obtained from standard flow measurement
methods~\cite{Voloshin:2008dg}. In addition, they feature a hierarchy
such that $v_{2} > v_{3} > v_{4}$ at all centralities except in the 5\% most 
central Pb--Pb collisions where the third is slightly larger
than the second harmonic, thereby explaining the presence of the
away-side double-hump seen in Fig.~\ref{fig:Figure1}. This is at
variance with the dependence of the regular flow coefficients which, even
in the centrality range 0--5\%, exhibit the basic hierarchy $v_{2} >
v_{3} > v_{4}$. The observed higher value of $v_{3}[P_2]$ relative
to $v_{2}[P_2]$ implies that $v_{3}$ should rise faster with
increasing $\pt $ than $v_2$, in agreement with explicit
measurements of the flow coefficients dependence on 
$\pt $~\cite{ALICE-PRL2011}.  

\begin{figure}[ht]
\begin{center}
  \includegraphics[scale=0.65]{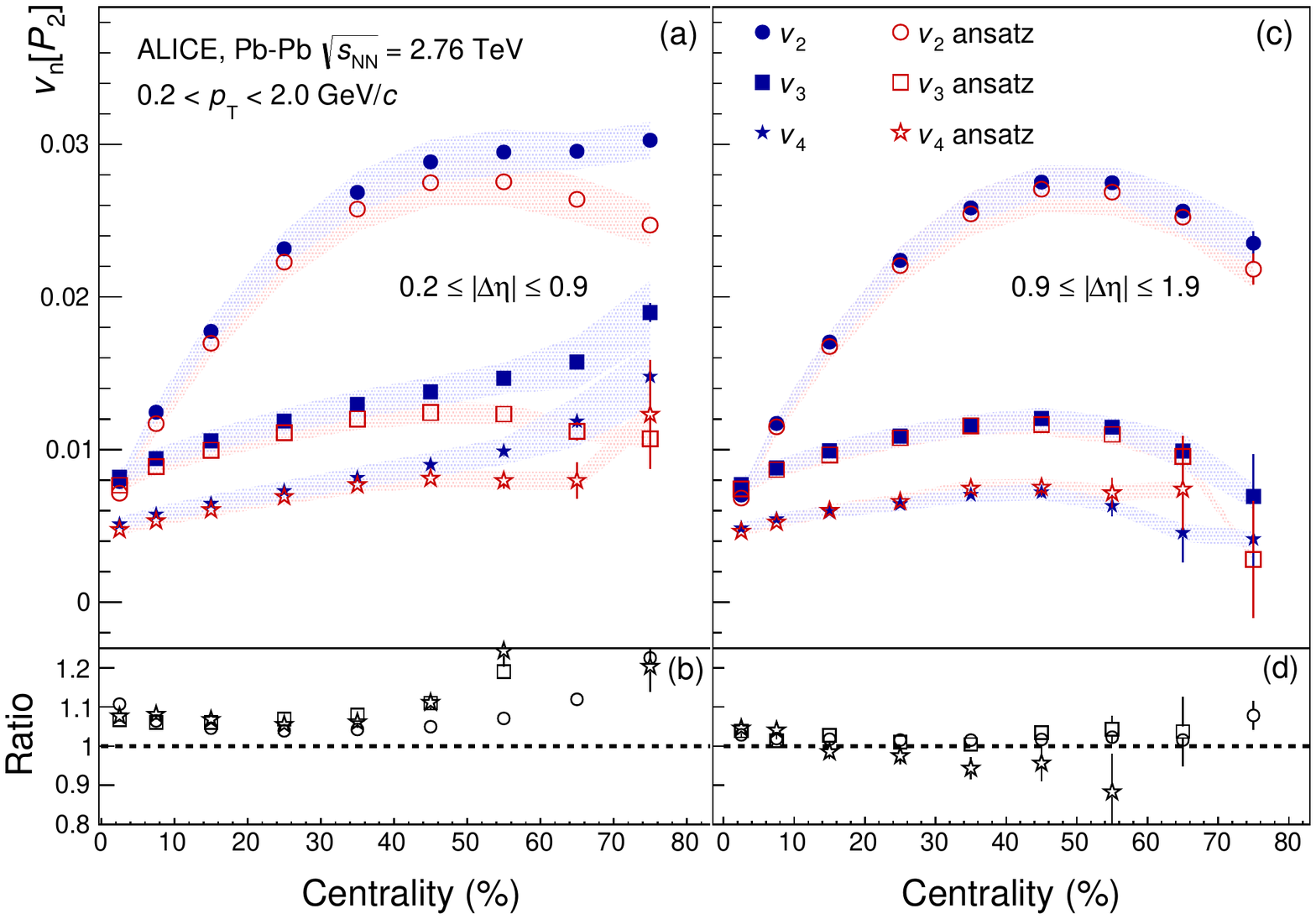}
  \caption{(Color online) $v_{n}$ coefficients where $n=2,3,4$ in the
    range (a) $0.2 \leq |\Delta\eta| \leq 0.9$ and 
    (c) $0.9 \leq |\Delta\eta| \leq 1.9$
    obtained from the $P_2$ correlation function. The coefficients are
    compared with the expectations from the flow ansatz calculated in
    their respective $\Deta$ ranges in Pb--Pb collisions. Statistical 
    errors are shown as vertical solid lines, whereas systematic
    errors are displayed as colored bands.
    Ratios of the $v_{n}$ coefficients and their 
    corresponding flow ansatz values are shown in (b) and (d). The errors on the
    ratios are only statistical.}
  \label{fig:Figure2}
\end{center}
\end{figure}
We next consider the possible role of non-flow correlations on the
correlator $P_2$ by comparing, in Fig. 2, the $v_{n}[P_2]$
coefficients obtained in ranges $0.2\le |\Delta\eta| \le 0.9$ 
and $0.9 \le |\Delta\eta| \le 1.9$ with values predicted by the flow 
ansatz, introduced above.  
In the range $0.9 \leq |\Delta\eta| \leq 1.9$ (see Fig. 2(c)), 
one observes that the coefficients
$v_n[P_2]$ are in a very good agreement, at all measured collision
centralities, with expectations from the flow ansatz. This agreement
provides additional evidence that two-particle correlations in this
relative pseudorapidity range are predominantly determined by the
collective nature of particle emission at low $\pt $, which motivates 
the factorization hypothesis used to derive Eq. (2). It also suggests that away-side jets, that might be associated with 
the near-side peak, are significantly suppressed and contribute minimally to the away-side correlated yield in that $\eta$ range.  
In contrast, in the range $0.2 \leq |\Delta\eta| \leq 0.9$ 
(see Fig. 2(a)), the
$v_n[P_2]$ coefficients exhibit a stronger and monotonic centrality
evolution. In particular, the $v_n[P_2]$
deviate  significantly from the flow ansatz for collision
centralities larger than 40\%, where one expects the largest non-flow
contributions associated with the presence of the correlation function
near-side peak.  

Using the same measurement technique, we further compare features of
the $P_2$ correlation function to that of the number correlation
function, $R_2$, defined as
\newline
\begin{equation}
R_2 +1 = 
\int_{p_{\rm T,min}}^{p_{\rm T,max}} ~ \rhoTwo(\pvecOne,\pvecTwo)\, ~
     \dptOne \dptTwo \, /
     \int_{p_{\rm T,min}}^{p_{\rm T,max}} {\rho_1(\pvecOne)\, \rho_1(\pvecTwo) ~
         \dptOne \dptTwo}.
\end{equation} 

\begin{figure}[ht]
\begin{center}
  \includegraphics[scale=0.66]{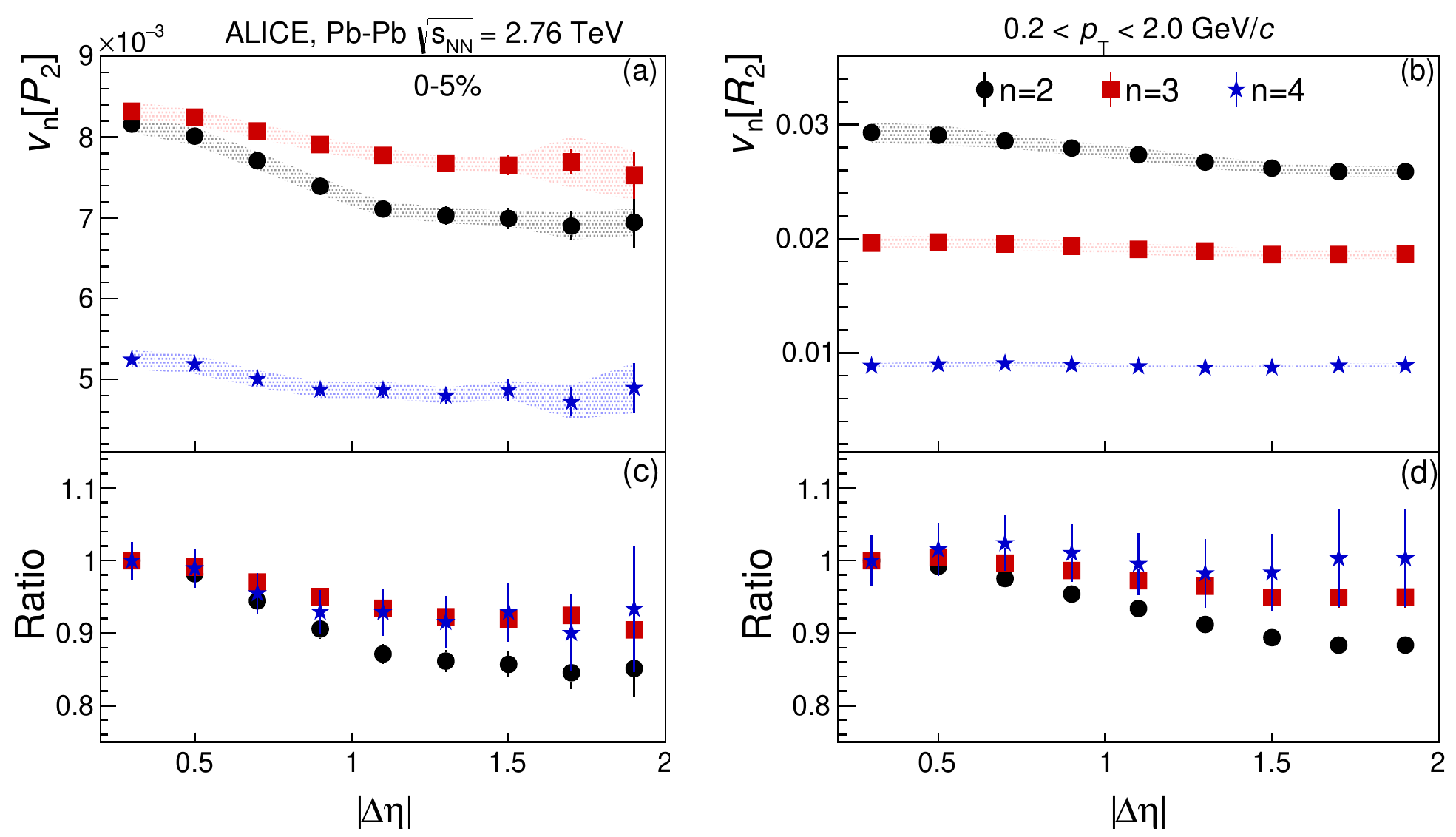}
\caption{(Color online) $v_{n}$ coefficients,  $n=2, 3, 4$, obtained 
  from (a) $P_2$ and (b) $R_2$ correlators,
  as a function of $|\Delta\eta|$ in the 5\% most central Pb -- Pb collisions.
  Statistical  errors are shown as vertical solid lines whereas  systematic
  uncertainties are displayed as shaded bands. Panels (c -- d): ratios of the $v_{n}$,  $n=2, 3, 4$,
  by the corresponding values of $v_{n}$ measured at $\Delta\eta = 0.3$.}
\label{fig:Figure3}
\end{center}
\end{figure}

Figure ~\ref{fig:Figure3} presents the $\Delta\eta$ dependence of
$v_{n}$, $n=$2, 3 and 4, coefficients obtained from these correlation
functions for the 5\% most central collisions. In this centrality interval 
one finds that the
hierarchies $v_{3}[P_2] > v_{2}[P_2]$ and $v_{2}[R_2] > v_{3}[R_2]$
indeed hold for all measured $\Delta\eta$. 
The dominance of $v_{3}[P_2]$ across all $\Delta\eta$
is likely a consequence of  the  third harmonic's (triangular flow) 
stronger dependence on $\pt $ relative to that of the second harmonic (elliptic 
flow).  The $v_2$, $v_3$, and $v_4$
dependencies on $\Delta\eta$ reveal additional interesting
features. In the case of the $R_2$ correlation, the coefficients $v_2$ and $v_3$ monotonically decrease over the entire  $\Delta\eta$ range, whereas coefficients extracted from $P_2$ exhibit a more pronounced decrease for $|\Delta\eta| \le  0.9$. From $|\Delta\eta| \sim 1.0$ to $\sim 2.0$, the relative decrease of $v_2$ is about 5\% for both correlators, and somewhat smaller for $v_3$.
These contrasting dependencies reflect the different shapes of
the near-side peaks of the two correlation functions. The narrower
shape of the near-side peak of the $P_2$ distribution 
suggests that the near-side peak of
$R_2$ might involve two components, one of which is characterized by
a vanishing $\rhoDptDpt$ for pairs with
$|\Delta\eta| \le 0.9$. While the origin of this behavior is not fully
understood, it offers the benefit of enabling the determination of
flow coefficients with smaller non-flow effects using a narrower
$\Delta\eta$ gap. 

In summary, we presented the first measurements of the 
two-particle transverse momentum differential
correlation function $P_2$
from Pb--Pb collisions at $\sqrt{s_{\rm NN}} =$ 2.76 TeV. In the 5\% most
central Pb--Pb collisions, $P_2$ has a shape qualitatively different
to that observed in measurements of the number density correlations,
with a relatively narrow near-side peak near $|\Deta|, \; |\Dphi| <
0.5$, and a longitudinally broad and double-hump structure on the
away-side. The double-hump structure in the 5\% most central 
$P_2$ correlation
indicates that this observable is more sensitive to the presence of a
triangular flow component than the number correlations $R_2$, and
consequently provides an indication that triangular flow features a
stronger dependence on $\pt $ than elliptic flow does.  
Comparison of the Fourier decompositions of the $R_2$ and $P_2$ 
correlators, calculated as a function of $|\Delta\eta|$, 
suggests that the  $v_2$, $v_3$, and $v_4$ coefficients extracted from  
$P_2$  reach approximately constant values beyond $|\Delta\eta| \sim 0.9$, 
while coefficients $v_2$, $v_3$ obtained from $R_2$  decrease 
monotonically for increasing $|\Delta\eta|$.
The observed agreement  between the flow coefficients measured from $P_2$
correlations, at $|\Delta\eta| > 0.9$, and the values predicted from 
the flow ansatz provide new and independent support to 
the notion that the observed long range correlations are largely due to
the initial collision geometry. These results may be used to
further constrain particle production models. This agreement to the
flow ansatz also provides further evidence for flow coefficient
factorization in heavy-ion collisions.

\newenvironment{acknowledgement}{\relax}{\relax}
\begin{acknowledgement}
\section*{Acknowledgements}

The ALICE Collaboration would like to thank all its engineers and technicians for their invaluable contributions to the construction of the experiment and the CERN accelerator teams for the outstanding performance of the LHC complex.
The ALICE Collaboration gratefully acknowledges the resources and support provided by all Grid centres and the Worldwide LHC Computing Grid (WLCG) collaboration.
The ALICE Collaboration acknowledges the following funding agencies for their support in building and running the ALICE detector:
A. I. Alikhanyan National Science Laboratory (Yerevan Physics Institute) Foundation (ANSL), State Committee of Science and World Federation of Scientists (WFS), Armenia;
Austrian Academy of Sciences and Nationalstiftung f\"{u}r Forschung, Technologie und Entwicklung, Austria;
Conselho Nacional de Desenvolvimento Cient\'{\i}fico e Tecnol\'{o}gico (CNPq), Universidade Federal do Rio Grande do Sul (UFRGS), Financiadora de Estudos e Projetos (Finep) and Funda\c{c}\~{a}o de Amparo \`{a} Pesquisa do Estado de S\~{a}o Paulo (FAPESP), Brazil;
Ministry of Science \& Technology of China (MSTC), National Natural Science Foundation of China (NSFC) and Ministry of Education of China (MOEC) , China;
Ministry of Science, Education and Sport and Croatian Science Foundation, Croatia;
Ministry of Education, Youth and Sports of the Czech Republic, Czech Republic;
The Danish Council for Independent Research | Natural Sciences, the Carlsberg Foundation and Danish National Research Foundation (DNRF), Denmark;
Helsinki Institute of Physics (HIP), Finland;
Commissariat \`{a} l'Energie Atomique (CEA) and Institut National de Physique Nucl\'{e}aire et de Physique des Particules (IN2P3) and Centre National de la Recherche Scientifique (CNRS), France;
Bundesministerium f\"{u}r Bildung, Wissenschaft, Forschung und Technologie (BMBF) and GSI Helmholtzzentrum f\"{u}r Schwerionenforschung GmbH, Germany;
Ministry of Education, Research and Religious Affairs, Greece;
National Research, Development and Innovation Office, Hungary;
Department of Atomic Energy Government of India (DAE) and Council of Scientific and Industrial Research (CSIR), New Delhi, India;
Indonesian Institute of Science, Indonesia;
Centro Fermi - Museo Storico della Fisica e Centro Studi e Ricerche Enrico Fermi and Istituto Nazionale di Fisica Nucleare (INFN), Italy;
Institute for Innovative Science and Technology , Nagasaki Institute of Applied Science (IIST), Japan Society for the Promotion of Science (JSPS) KAKENHI and Japanese Ministry of Education, Culture, Sports, Science and Technology (MEXT), Japan;
Consejo Nacional de Ciencia (CONACYT) y Tecnolog\'{i}a, through Fondo de Cooperaci\'{o}n Internacional en Ciencia y Tecnolog\'{i}a (FONCICYT) and Direcci\'{o}n General de Asuntos del Personal Academico (DGAPA), Mexico;
Nationaal instituut voor subatomaire fysica (Nikhef), Netherlands;
The Research Council of Norway, Norway;
Commission on Science and Technology for Sustainable Development in the South (COMSATS), Pakistan;
Pontificia Universidad Cat\'{o}lica del Per\'{u}, Peru;
Ministry of Science and Higher Education and National Science Centre, Poland;
Korea Institute of Science and Technology Information and National Research Foundation of Korea (NRF), Republic of Korea;
Ministry of Education and Scientific Research, Institute of Atomic Physics and Romanian National Agency for Science, Technology and Innovation, Romania;
Joint Institute for Nuclear Research (JINR), Ministry of Education and Science of the Russian Federation and National Research Centre Kurchatov Institute, Russia;
Ministry of Education, Science, Research and Sport of the Slovak Republic, Slovakia;
National Research Foundation of South Africa, South Africa;
Centro de Aplicaciones Tecnol\'{o}gicas y Desarrollo Nuclear (CEADEN), Cubaenerg\'{\i}a, Cuba, Ministerio de Ciencia e Innovacion and Centro de Investigaciones Energ\'{e}ticas, Medioambientales y Tecnol\'{o}gicas (CIEMAT), Spain;
Swedish Research Council (VR) and Knut \& Alice Wallenberg Foundation (KAW), Sweden;
European Organization for Nuclear Research, Switzerland;
National Science and Technology Development Agency (NSDTA), Suranaree University of Technology (SUT) and Office of the Higher Education Commission under NRU project of Thailand, Thailand;
Turkish Atomic Energy Agency (TAEK), Turkey;
National Academy of  Sciences of Ukraine, Ukraine;
Science and Technology Facilities Council (STFC), United Kingdom;
National Science Foundation of the United States of America (NSF) and United States Department of Energy, Office of Nuclear Physics (DOE NP), United States of America.    
\end{acknowledgement}

\bibliography{dptdpt-prp-V11}
\bibliographystyle{utphys}   

\newpage
\appendix
\section{The ALICE Collaboration}
\label{app:collab}



\begingroup
\small
\begin{flushleft}
J.~Adam$^\textrm{\scriptsize 38}$,
D.~Adamov\'{a}$^\textrm{\scriptsize 87}$,
M.M.~Aggarwal$^\textrm{\scriptsize 91}$,
G.~Aglieri Rinella$^\textrm{\scriptsize 34}$,
M.~Agnello$^\textrm{\scriptsize 30}$\textsuperscript{,}$^\textrm{\scriptsize 113}$,
N.~Agrawal$^\textrm{\scriptsize 47}$,
Z.~Ahammed$^\textrm{\scriptsize 137}$,
S.~Ahmad$^\textrm{\scriptsize 17}$,
S.U.~Ahn$^\textrm{\scriptsize 70}$,
S.~Aiola$^\textrm{\scriptsize 141}$
% ...
\renewcommand\labelenumi{\textsuperscript{\theenumi}~}

\section*{Affiliation notes}
\renewcommand\theenumi{\roman{enumi}}
\begin{Authlist}
\item \Adef{0}Deceased
\item \Adef{idp1753616}{Also at: Georgia State University, Atlanta, Georgia, United States}
\item \Adef{idp3184896}{Also at: Also at Department of Applied Physics, Aligarh Muslim University, Aligarh, India}
\item \Adef{idp3956384}{Also at: M.V. Lomonosov Moscow State University, D.V. Skobeltsyn Institute of Nuclear, Physics, Moscow, Russia}
\end{Authlist}

\section*{Collaboration Institutes}
\renewcommand\theenumi{\arabic{enumi}~}

$^{1}$A.I. Alikhanyan National Science Laboratory (Yerevan Physics Institute) Foundation, Yerevan, Armenia
\\
$^{2}$Benem\'{e}rita Universidad Aut\'{o}noma de Puebla, Puebla, Mexico
\\
$^{3}$Bogolyubov Institute for Theoretical Physics, Kiev, Ukraine
\\
$^{4}$Bose Institute, Department of Physics 
and Centre for Astroparticle Physics and Space Science (CAPSS), Kolkata, India
\\
$^{5}$Budker Institute for Nuclear Physics, Novosibirsk, Russia
\\
$^{6}$California Polytechnic State University, San Luis Obispo, California, United States
\\
$^{7}$Central China Normal University, Wuhan, China
\\
$^{8}$Centre de Calcul de l'IN2P3, Villeurbanne, Lyon, France
\\
$^{9}$Centro de Aplicaciones Tecnol\'{o}gicas y Desarrollo Nuclear (CEADEN), Havana, Cuba
\\
$^{10}$Centro de Investigaciones Energ\'{e}ticas Medioambientales y Tecnol\'{o}gicas (CIEMAT), Madrid, Spain
\\
$^{11}$Centro de Investigaci\'{o}n y de Estudios Avanzados (CINVESTAV), Mexico City and M\'{e}rida, Mexico
\\
$^{12}$Centro Fermi - Museo Storico della Fisica e Centro Studi e Ricerche ``Enrico Fermi', Rome, Italy
\\
$^{13}$Chicago State University, Chicago, Illinois, United States
\\
$^{14}$China Institute of Atomic Energy, Beijing, China
\\
$^{15}$COMSATS Institute of Information Technology (CIIT), Islamabad, Pakistan
\\
$^{16}$Departamento de F\'{\i}sica de Part\'{\i}culas and IGFAE, Universidad de Santiago de Compostela, Santiago de Compostela, Spain
\\
$^{17}$Department of Physics, Aligarh Muslim University, Aligarh, India
\\
$^{18}$Department of Physics, Ohio State University, Columbus, Ohio, United States
\\
$^{19}$Department of Physics, Sejong University, Seoul, South Korea
\\
$^{20}$Department of Physics, University of Oslo, Oslo, Norway
\\
$^{21}$Department of Physics and Technology, University of Bergen, Bergen, Norway
\\
$^{22}$Dipartimento di Fisica dell'Universit\`{a} 'La Sapienza'
and Sezione INFN, Rome, Italy
\\
$^{23}$Dipartimento di Fisica dell'Universit\`{a}
and Sezione INFN, Cagliari, Italy
\\
$^{24}$Dipartimento di Fisica dell'Universit\`{a}
and Sezione INFN, Trieste, Italy
\\
$^{25}$Dipartimento di Fisica dell'Universit\`{a}
and Sezione INFN, Turin, Italy
\\
$^{26}$Dipartimento di Fisica e Astronomia dell'Universit\`{a}
and Sezione INFN, Bologna, Italy
\\
$^{27}$Dipartimento di Fisica e Astronomia dell'Universit\`{a}
and Sezione INFN, Catania, Italy
\\
$^{28}$Dipartimento di Fisica e Astronomia dell'Universit\`{a}
and Sezione INFN, Padova, Italy
\\
$^{29}$Dipartimento di Fisica `E.R.~Caianiello' dell'Universit\`{a}
and Gruppo Collegato INFN, Salerno, Italy
\\
$^{30}$Dipartimento DISAT del Politecnico and Sezione INFN, Turin, Italy
\\
$^{31}$Dipartimento di Scienze e Innovazione Tecnologica dell'Universit\`{a} del Piemonte Orientale and INFN Sezione di Torino, Alessandria, Italy
\\
$^{32}$Dipartimento Interateneo di Fisica `M.~Merlin'
and Sezione INFN, Bari, Italy
\\
$^{33}$Division of Experimental High Energy Physics, University of Lund, Lund, Sweden
\\
$^{34}$European Organization for Nuclear Research (CERN), Geneva, Switzerland
\\
$^{35}$Excellence Cluster Universe, Technische Universit\"{a}t M\"{u}nchen, Munich, Germany
\\
$^{36}$Faculty of Engineering, Bergen University College, Bergen, Norway
\\
$^{37}$Faculty of Mathematics, Physics and Informatics, Comenius University, Bratislava, Slovakia
\\
$^{38}$Faculty of Nuclear Sciences and Physical Engineering, Czech Technical University in Prague, Prague, Czech Republic
\\
$^{39}$Faculty of Science, P.J.~\v{S}af\'{a}rik University, Ko\v{s}ice, Slovakia
\\
$^{40}$Faculty of Technology, Buskerud and Vestfold University College, Tonsberg, Norway
\\
$^{41}$Frankfurt Institute for Advanced Studies, Johann Wolfgang Goethe-Universit\"{a}t Frankfurt, Frankfurt, Germany
\\
$^{42}$Gangneung-Wonju National University, Gangneung, South Korea
\\
$^{43}$Gauhati University, Department of Physics, Guwahati, India
\\
$^{44}$Helmholtz-Institut f\"{u}r Strahlen- und Kernphysik, Rheinische Friedrich-Wilhelms-Universit\"{a}t Bonn, Bonn, Germany
\\
$^{45}$Helsinki Institute of Physics (HIP), Helsinki, Finland
\\
$^{46}$Hiroshima University, Hiroshima, Japan
\\
$^{47}$Indian Institute of Technology Bombay (IIT), Mumbai, India
\\
$^{48}$Indian Institute of Technology Indore, Indore, India
\\
$^{49}$Indonesian Institute of Sciences, Jakarta, Indonesia
\\
$^{50}$Inha University, Incheon, South Korea
\\
$^{51}$Institut de Physique Nucl\'eaire d'Orsay (IPNO), Universit\'e Paris-Sud, CNRS-IN2P3, Orsay, France
\\
$^{52}$Institute for Nuclear Research, Academy of Sciences, Moscow, Russia
\\
$^{53}$Institute for Subatomic Physics of Utrecht University, Utrecht, Netherlands
\\
$^{54}$Institute for Theoretical and Experimental Physics, Moscow, Russia
\\
$^{55}$Institute of Experimental Physics, Slovak Academy of Sciences, Ko\v{s}ice, Slovakia
\\
$^{56}$Institute of Physics, Academy of Sciences of the Czech Republic, Prague, Czech Republic
\\
$^{57}$Institute of Physics, Bhubaneswar, India
\\
$^{58}$Institute of Space Science (ISS), Bucharest, Romania
\\
$^{59}$Institut f\"{u}r Informatik, Johann Wolfgang Goethe-Universit\"{a}t Frankfurt, Frankfurt, Germany
\\
$^{60}$Institut f\"{u}r Kernphysik, Johann Wolfgang Goethe-Universit\"{a}t Frankfurt, Frankfurt, Germany
\\
$^{61}$Institut f\"{u}r Kernphysik, Westf\"{a}lische Wilhelms-Universit\"{a}t M\"{u}nster, M\"{u}nster, Germany
\\
$^{62}$Instituto de Ciencias Nucleares, Universidad Nacional Aut\'{o}noma de M\'{e}xico, Mexico City, Mexico
\\
$^{63}$Instituto de F\'{i}sica, Universidade Federal do Rio Grande do Sul (UFRGS), Porto Alegre, Brazil
\\
$^{64}$Instituto de F\'{\i}sica, Universidad Nacional Aut\'{o}noma de M\'{e}xico, Mexico City, Mexico
\\
$^{65}$Institut Pluridisciplinaire Hubert Curien (IPHC), Universit\'{e} de Strasbourg, CNRS-IN2P3, Strasbourg, France
\\
$^{66}$IRFU, CEA, Universit\'{e} Paris-Saclay, F-91191 Gif-sur-Yvette, France, Saclay, France
\\
$^{67}$iThemba LABS, National Research Foundation, Somerset West, South Africa
\\
$^{68}$Joint Institute for Nuclear Research (JINR), Dubna, Russia
\\
$^{69}$Konkuk University, Seoul, South Korea
\\
$^{70}$Korea Institute of Science and Technology Information, Daejeon, South Korea
\\
$^{71}$KTO Karatay University, Konya, Turkey
\\
$^{72}$Laboratoire de Physique Corpusculaire (LPC), Clermont Universit\'{e}, Universit\'{e} Blaise Pascal, CNRS--IN2P3, Clermont-Ferrand, France
\\
$^{73}$Laboratoire de Physique Subatomique et de Cosmologie, Universit\'{e} Grenoble-Alpes, CNRS-IN2P3, Grenoble, France
\\
$^{74}$Laboratori Nazionali di Frascati, INFN, Frascati, Italy
\\
$^{75}$Laboratori Nazionali di Legnaro, INFN, Legnaro, Italy
\\
$^{76}$Lawrence Berkeley National Laboratory, Berkeley, California, United States
\\
$^{77}$Moscow Engineering Physics Institute, Moscow, Russia
\\
$^{78}$Nagasaki Institute of Applied Science, Nagasaki, Japan
\\
$^{79}$National and Kapodistrian University of Athens, Physics Department, Athens, Greece, Athens, Greece
\\
$^{80}$National Centre for Nuclear Studies, Warsaw, Poland
\\
$^{81}$National Institute for Physics and Nuclear Engineering, Bucharest, Romania
\\
$^{82}$National Institute of Science Education and Research, Bhubaneswar, India
\\
$^{83}$National Research Centre Kurchatov Institute, Moscow, Russia
\\
$^{84}$Niels Bohr Institute, University of Copenhagen, Copenhagen, Denmark
\\
$^{85}$Nikhef, Nationaal instituut voor subatomaire fysica, Amsterdam, Netherlands
\\
$^{86}$Nuclear Physics Group, STFC Daresbury Laboratory, Daresbury, United Kingdom
\\
$^{87}$Nuclear Physics Institute, Academy of Sciences of the Czech Republic, \v{R}e\v{z} u Prahy, Czech Republic
\\
$^{88}$Oak Ridge National Laboratory, Oak Ridge, Tennessee, United States
\\
$^{89}$Petersburg Nuclear Physics Institute, Gatchina, Russia
\\
$^{90}$Physics Department, Creighton University, Omaha, Nebraska, United States
\\
$^{91}$Physics Department, Panjab University, Chandigarh, India
\\
$^{92}$Physics Department, University of Cape Town, Cape Town, South Africa
\\
$^{93}$Physics Department, University of Jammu, Jammu, India
\\
$^{94}$Physics Department, University of Rajasthan, Jaipur, India
\\
$^{95}$Physikalisches Institut, Eberhard Karls Universit\"{a}t T\"{u}bingen, T\"{u}bingen, Germany
\\
$^{96}$Physikalisches Institut, Ruprecht-Karls-Universit\"{a}t Heidelberg, Heidelberg, Germany
\\
$^{97}$Physik Department, Technische Universit\"{a}t M\"{u}nchen, Munich, Germany
\\
$^{98}$Purdue University, West Lafayette, Indiana, United States
\\
$^{99}$Pusan National University, Pusan, South Korea
\\
$^{100}$Research Division and ExtreMe Matter Institute EMMI, GSI Helmholtzzentrum f\"ur Schwerionenforschung GmbH, Darmstadt, Germany
\\
$^{101}$Rudjer Bo\v{s}kovi\'{c} Institute, Zagreb, Croatia
\\
$^{102}$Russian Federal Nuclear Center (VNIIEF), Sarov, Russia
\\
$^{103}$Saha Institute of Nuclear Physics, Kolkata, India
\\
$^{104}$School of Physics and Astronomy, University of Birmingham, Birmingham, United Kingdom
\\
$^{105}$Secci\'{o}n F\'{\i}sica, Departamento de Ciencias, Pontificia Universidad Cat\'{o}lica del Per\'{u}, Lima, Peru
\\
$^{106}$Sezione INFN, Bari, Italy
\\
$^{107}$Sezione INFN, Bologna, Italy
\\
$^{108}$Sezione INFN, Cagliari, Italy
\\
$^{109}$Sezione INFN, Catania, Italy
\\
$^{110}$Sezione INFN, Padova, Italy
\\
$^{111}$Sezione INFN, Rome, Italy
\\
$^{112}$Sezione INFN, Trieste, Italy
\\
$^{113}$Sezione INFN, Turin, Italy
\\
$^{114}$SSC IHEP of NRC Kurchatov institute, Protvino, Russia
\\
$^{115}$Stefan Meyer Institut f\"{u}r Subatomare Physik (SMI), Vienna, Austria
\\
$^{116}$SUBATECH, Ecole des Mines de Nantes, Universit\'{e} de Nantes, CNRS-IN2P3, Nantes, France
\\
$^{117}$Suranaree University of Technology, Nakhon Ratchasima, Thailand
\\
$^{118}$Technical University of Ko\v{s}ice, Ko\v{s}ice, Slovakia
\\
$^{119}$Technical University of Split FESB, Split, Croatia
\\
$^{120}$The Henryk Niewodniczanski Institute of Nuclear Physics, Polish Academy of Sciences, Cracow, Poland
\\
$^{121}$The University of Texas at Austin, Physics Department, Austin, Texas, United States
\\
$^{122}$Universidad Aut\'{o}noma de Sinaloa, Culiac\'{a}n, Mexico
\\
$^{123}$Universidade de S\~{a}o Paulo (USP), S\~{a}o Paulo, Brazil
\\
$^{124}$Universidade Estadual de Campinas (UNICAMP), Campinas, Brazil
\\
$^{125}$Universidade Federal do ABC, Santo Andre, Brazil
\\
$^{126}$University of Houston, Houston, Texas, United States
\\
$^{127}$University of Jyv\"{a}skyl\"{a}, Jyv\"{a}skyl\"{a}, Finland
\\
$^{128}$University of Liverpool, Liverpool, United Kingdom
\\
$^{129}$University of Tennessee, Knoxville, Tennessee, United States
\\
$^{130}$University of the Witwatersrand, Johannesburg, South Africa
\\
$^{131}$University of Tokyo, Tokyo, Japan
\\
$^{132}$University of Tsukuba, Tsukuba, Japan
\\
$^{133}$University of Zagreb, Zagreb, Croatia
\\
$^{134}$Universit\'{e} de Lyon, Universit\'{e} Lyon 1, CNRS/IN2P3, IPN-Lyon, Villeurbanne, Lyon, France
\\
$^{135}$Universit\`{a} di Brescia, Brescia, Italy
\\
$^{136}$V.~Fock Institute for Physics, St. Petersburg State University, St. Petersburg, Russia
\\
$^{137}$Variable Energy Cyclotron Centre, Kolkata, India
\\
$^{138}$Warsaw University of Technology, Warsaw, Poland
\\
$^{139}$Wayne State University, Detroit, Michigan, United States
\\
$^{140}$Wigner Research Centre for Physics, Hungarian Academy of Sciences, Budapest, Hungary
\\
$^{141}$Yale University, New Haven, Connecticut, United States
\\
$^{142}$Yonsei University, Seoul, South Korea
\\
$^{143}$Zentrum f\"{u}r Technologietransfer und Telekommunikation (ZTT), Fachhochschule Worms, Worms, Germany
\endgroup

\end{document}